\documentclass{article}

\usepackage[letterpaper, top=1in, bottom=1.2in]{geometry}
\newif\ifcomment
\usepackage{authblk}

\usepackage{graphicx}
\usepackage{dcolumn} 
\usepackage{bm}      
\usepackage{amssymb} 
\usepackage{amsfonts}
\usepackage{graphics}
\usepackage{grffile} 
\usepackage{epsfig}
\usepackage[loose]{units}
\usepackage[usenames,dvipsnames]{xcolor}
\usepackage[normalem]{ulem}
\usepackage{color}
\usepackage{hyperref}
\usepackage{cleveref}
\begin{document}
\title{Amplitude-tuneable, octal LED pulser for emulation of scintillation light in multi-channel  photon detectors}
\author[1]{Alexandru Rusu}
\affil[1]{ORNL, Oak Ridge National Laboratory, TN37831, USA}
\author[2]{Hans Muller}
\affil[2]{CERN PH, CH1211 Geneva 23, Switzerland}
\date{\today}
\maketitle
\begin{abstract}
 Photon detectors like the ALICE Calorimeters are vulnerable to both optical and electrical crosstalk between detector channels belonging to the same particle event.
 In order to quantify the crosstalk generated by a single channel on its neighbours, the standard method used so far are calibrated electrical pulses in order to quantify the crosstalk between individual frontend preamplifier channels.
 However due to grounding imperfections, cross-couplings via bias voltage supplies and imperfect electromagnetic shielding, this method introduces biases and is not precise.
 A new approach is the use of emulated scintillation light over optical fibre bundles in order to effectively generate  scintillation light in clustered channels.
 This method is based on generation of quasi time-coherent light with similar time profiles as  scintillation light.
 Theoretically, and  based on linear superposition, pulsing a single channel with calibrated light would be sufficient to measure the crosstalk in the surrounding channels.
 However, non-linear effects in real detector frontends with energy measurements over cluster sizes like 3x3 can add up to significant errors.
 Non-linearity is for example due to preamplifier supply voltage drop in the distribution and GND return lines, in particular if power is provided over long cables.
 This can generate locally reduced or increased signal amplitudes depending on cumulative signal intensity.
 The presented octal light ALED~(Analogue LED) pulser prototype emulates scintillation light with individually tunable light amplitudes in 8 channels.
 It can be used for pulsing single channels, or for clustered channels with a predefined amplitude profile.
 Unbiased single channel crosstalk and non-linear side-effects can be determined for single and  multichannel situations.
\end{abstract}
\section{Introduction}
In the following we assume the use of Avalanche Photodiodes~(APDs),  however in general, any other type of photodetector, or different test wavelengths may be used.
The ALED method emulates coherent scintillation light with tunable intensities in each channels.
The 8-channel prototype is equipped with high-luminosity blue InGan LEDs of type HLMP-CB3B-UVODD from Avago.  
The common light pulse over eight fibres is triggerable via a common NIM trigger input at rates up 10 kHz.
Each channel generates light with a center wavelength of 465 nm, a risetime of less than 1~ns and an envelope of $ O(15{\rm ns})$.
A optical fiber connector allows to insert polymer fibres of 3mm diameter  such that they face the full light cone of the LED.
With a  minimum wavelength attenuation of the fibres  the range between 450 and 550nm, the light pulses can get transmitted over several meters for measurements which do not induce neither electrical nor optical crosstalk.
For testing arrays of APDs and pre-amplifiers~\cite{Badala:2008zzd} of the ALICE calorimeters with octal fiber bundles, custom fibre adapters fitting with the APD diodes have been designed and produced with a 3D printer.
For thermal stability, all APDs were mounted on a common copper bar, allowing to control and measure their common temperature.
This is particularly important due to the high temperature coefficient of APD gains of O(2\%/)/C.  
The described test setup was built to measure unbiased crosstalk measurements between the APD channels of the ALICE EMCal detector~\cite{Allen:2009aa}.
Arrays of 32 APDs–and-CSP preamplifier get powered and are read-out via 32-channel readout cards (FEE)~\cite{Muller:2006jr}.
By connecting the fiber-ends of the octal bundle to selected APDs one can study crosstalk and linearity.
The FEE electronics [8] features individual, 12 bit APD gain control, allowing to calibrate all channels to the same light response.
The linear chain of  APD/CSP preamplifiers with  a common charge gain of 1mV/fC, followed by  semi-gaussian shaper with of 200ns peaking time, covers  a  dynamic range from 50 MeV-200GeV. 
However, since the primary scintillation light is distributed over clustered channels, energy needs to added over 2x2 or 3x3 regions.
The linearity of the superposition of channels strongly depends on crosstalk and other non-linearities. 
For quantifying these effects with single-channel light pulsing, the peak amplitudes of semi-gaussian envelopes of the FEE electronics shapers were measured for neigboring channels and analysed using a digital oscilloscope.  
Analysis of multichannel behaviour using readout and analysis via the associated Data Acquisition system (DATE) is planned.

\begin{figure}[t!] 
 \centering 
 \includegraphics[width=0.6\linewidth]{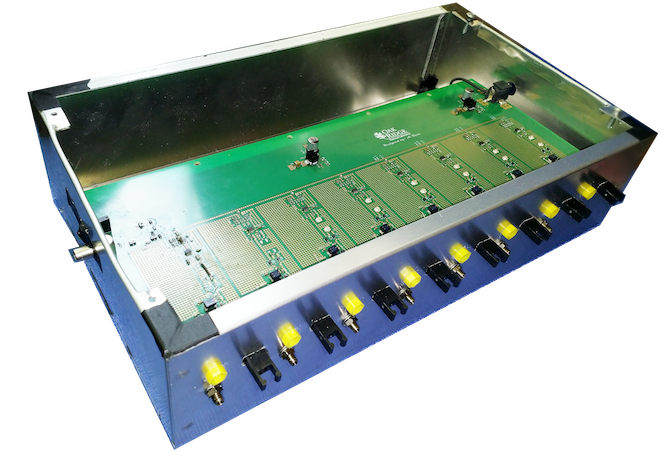}
 \caption{Octal-channel ALED box with fiber connectors.}
 \label{fig:1}
\end{figure}

The ALED pulser prototype (Fig.~\ref{fig:1}) was implemented on a single printed circuit board, fitting in a light-tight aluminium chassis.
The frontend side of the chassis presents eight snap-in optical connectors, each associated with an amplitude monitoring test-point and a 20-turn trimmer for the light intensity tuning of each channel.
Individual channels can get disabled via a slide switch. 
\section{LED pulser principles}
The ALED (Analogue LED) pulse principle is an alternative to a)~capacitor discharge method via avalanche transistors; b~) picosecond pulse generation via SRD diodes~\cite{protiva2010}.  
The avalanche transistor method was used for the EMCal LED monitoring system~\cite{Cormier:932676} however is limited to fixed light intensities.
The SRD diode technology provides picosecond-range pulses with possibility for variable intensities, however requires expensive components like microwave filters.  
ALED is a cost-effective alternative method, allowing for multichannel and coherent pulsing with $O(250{\rm ps})$ risetime and individual light intensities that can be tuned over a large dynamic range.  
\section{Fast pulse generation}
\begin{figure}[t!] 
 \centering 
 \includegraphics[width=0.95\linewidth]{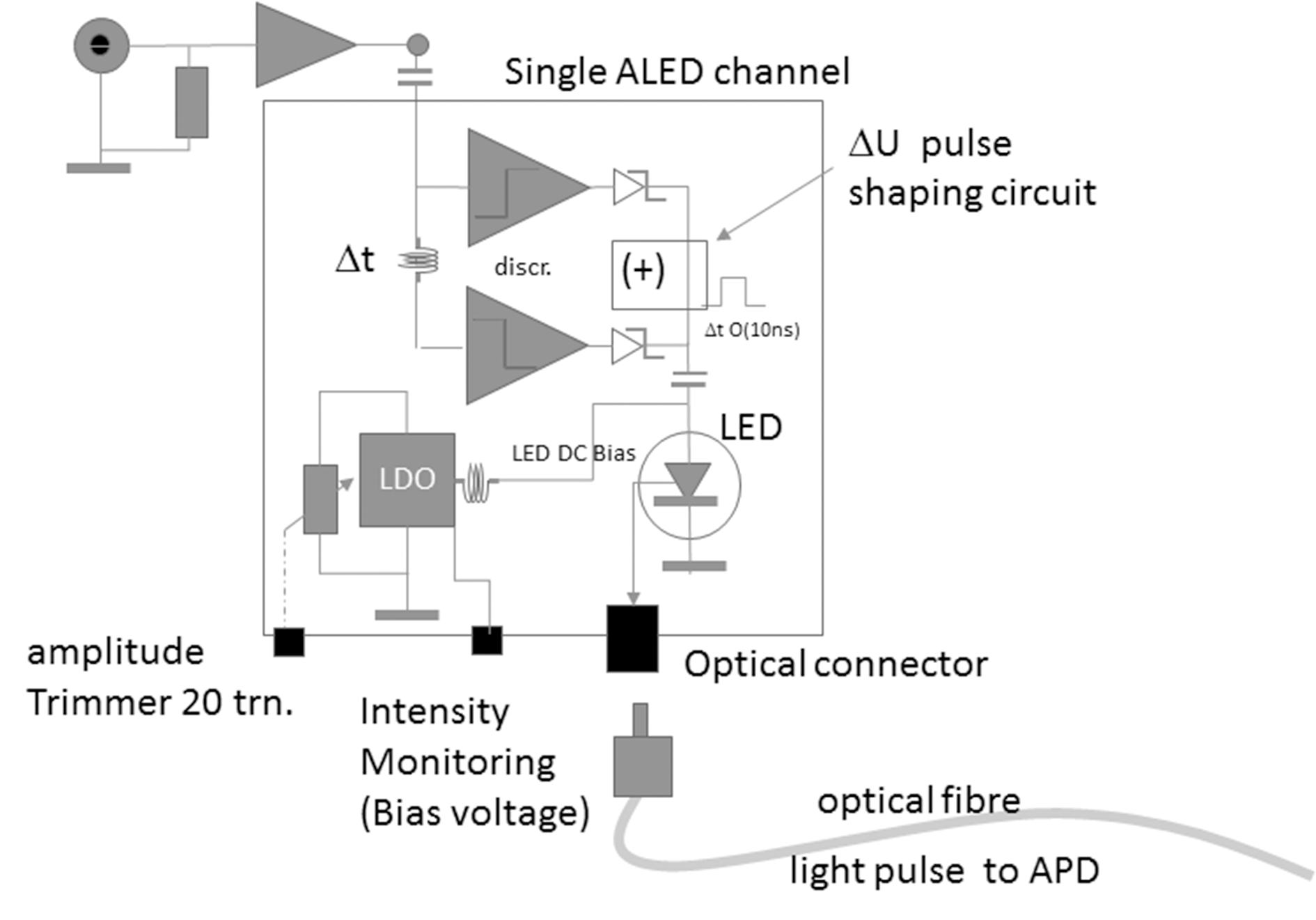}
 \caption{ALED block diagram (one channel).}
 \label{fig:2}
\end{figure}

Each ALED channel consists of two equal and very fast discriminator channels, triggered at slightly delayed times of order $\Delta t \sim 5$ns.
The block diagram is depicted in Fig.~\ref{fig:2}.
Unloaded output pulse amplitudes reach 5V with rise-times of $<1$ns.
We used dual discriminator chips of type TLV3502~\cite{ti}.
After adding their complementary outputs through a pulse shaping circuit, the prototype achieves risetims of $O(250{\rm ps})$ with 1.3V amplitude over  50$\Omega$.
The impedance of the LED diode is  dynamic,  starting at 500$\Omega$ below 1mA and settling to 10$\Omega$ above 30mA.
The low impedance at higher amplitudes leads to non-linear pulse reflections and signal distortion with effective envelope  prolongation at the falling pulse side.
The leading, $5$ns short rectangular pulse corresponds to the discriminator output.
The pulse shaping effects used in the ALED result in an effective LED light envelope of $O(1{\rm ns})$ risetime, a short high intensity pulse of $\Delta t = 5$ns,  followed by  a quasi exponentially  decaying intensity over $15$ns.
The $\Delta t=5$ns the high intensity leading pulse is defined by a delay circuit at the common input of the 2 discriminators. 
\section{Dynamic light range}
\begin{figure}[t!] 
  \centering 
 \includegraphics[width=0.75\linewidth]{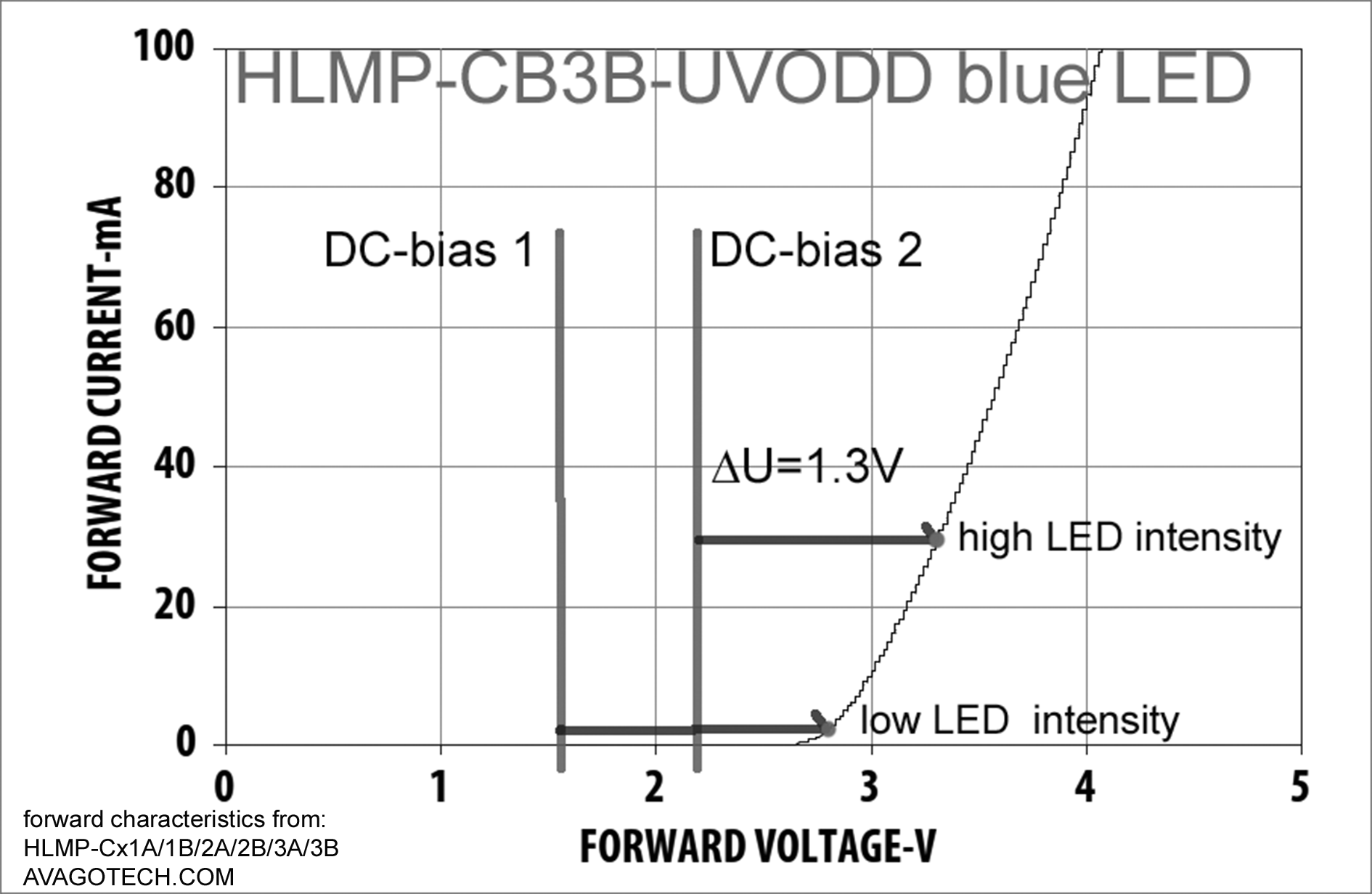}
 \caption{LED  forward light range.}
 \label{fig:3}
\end{figure}

The primary pulse generates LED light as an addition of the fixed amplitude discriminator output pulse of 1.3V  on top of a forward-biased point for the LED.
In absence of a triggered pulse, the LED produces no light at DC-bias equals 1V.
For a blue InGan LED of type HLMP-CB3B-UVODD the inset of light generation is at ca.\ 2.2V, well below the normally specified forward voltage 3V@20mA  depicted in Fig.~\ref{fig:3}. 
When the DC-bias in increased and adds up with the 1.3V pulse to more than 2.2V the LED is spontaneously driven into light emitting mode.
The intensity of the pulsed light can thus be precisely adjusted via the DC-bias level via a 20-turn potentiometer. 
The intensity range between the maximum and miniumum
DC-bias ideally corresponds to a LED pulse current from
0mA to 30mA. Practically, for the tests with the FEE
electronics  we have set the lowest light level at
DC-bias-1 = 1.5V and the highest at DC-bias = 2.2V. The
light intensity at 20mA@3.3V is specified for the given
LED (HLMP-CB3B-UVODD) as 3.2–5.5 candela, hence with
a 30mA pulse and an exponential increase we can
approximate a maximum luminous intensity of 10 lumen
per steradian for pulses with the maximum DC bias
setting. 
The LED pulse measurement  with the EMCal FEE electronics chain confirms that the full dynamic detector range of 14 bit is covered by this choice of LED and its bias settings.
A factor 2 in intensity would is available by choosing LEDs of the same type but higher luminosity (HLMP-CB1B-XY0DD ).
\begin{figure}[t!] 
  \centering 
 \includegraphics[width=0.75\linewidth]{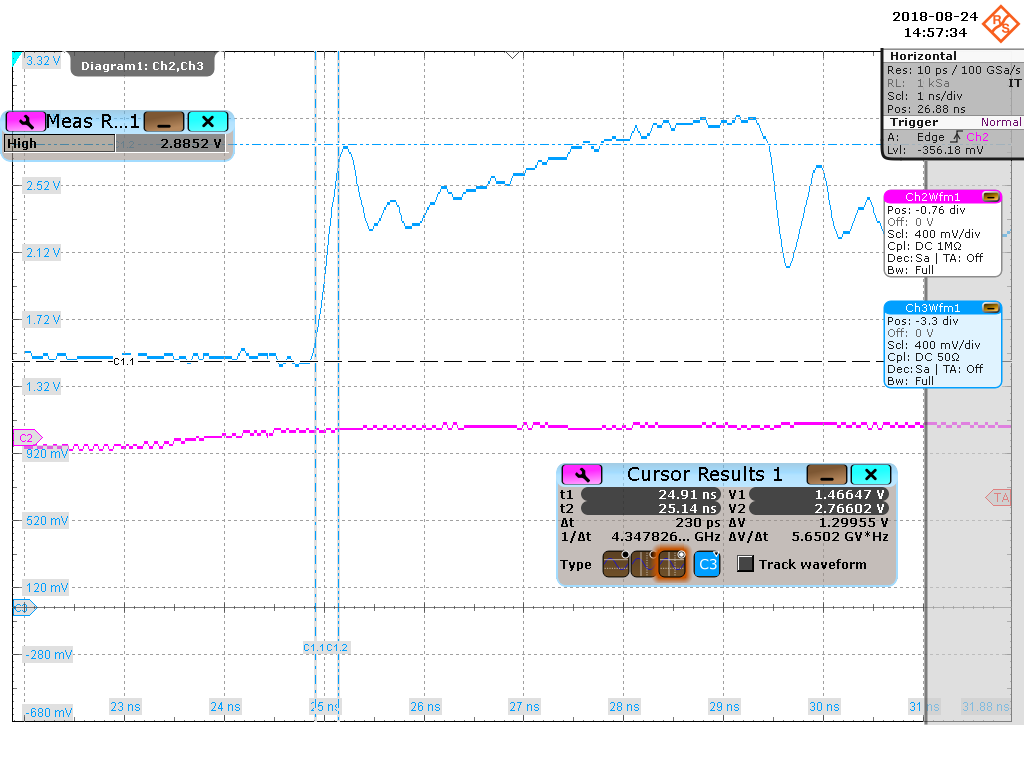}
 \caption{leading LED pulse with fast risetime.}
 \label{fig:4}
\end{figure}

\section{Light pulse timing profile}
\begin{figure}[t!] 
  \centering 
 \includegraphics[width=0.75\linewidth]{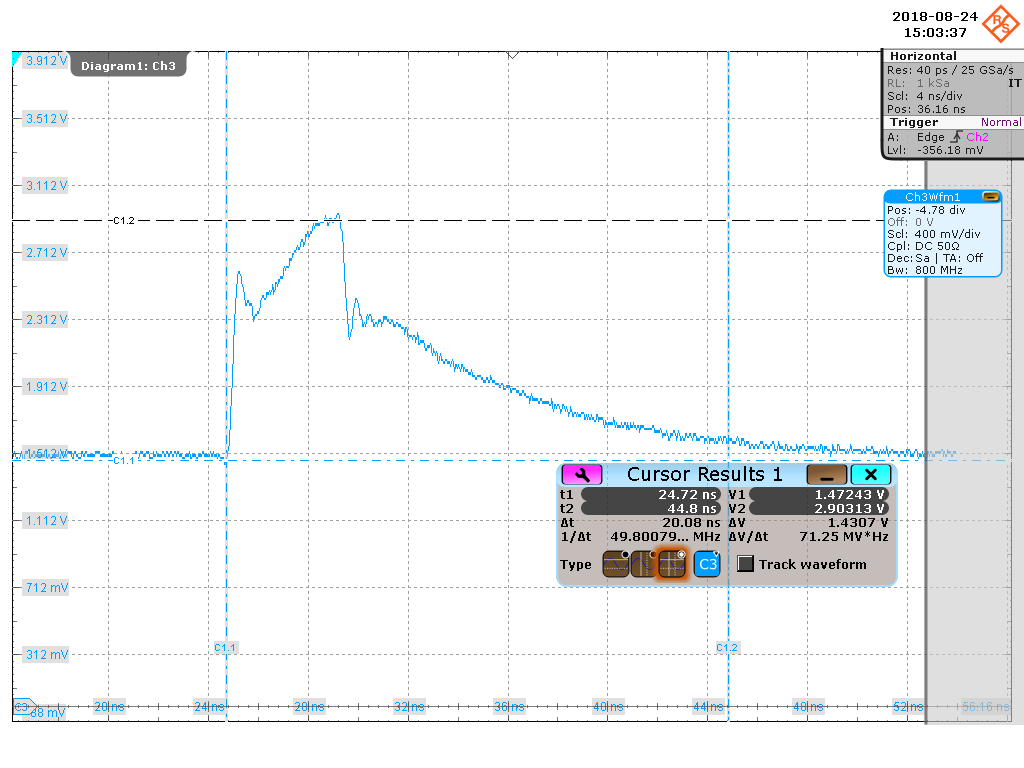}
 \caption{LED pulse envelope with decaying trailer.}
 \label{fig:5}
\end{figure}

The effective LED pulse was measured using a 3GHz, 50$\Omega$  probe across the LED.
The pulse envelope (Fig.~\ref{fig:5}) consists of a leading pulse with 10\%--90\% risetime of 250ps and 1.3V peak amplitude (Fig.~\ref{fig:4}) followed by reflections due to the non-linear LED impedance 
The resulting light envelope has a similar time profile as the scintillation light generated by scintillating fibers.
o
\section{Test of electronic chain with LED pulser}
The readout electronics chain of EMCal~\cite{Muller:2006jr} detects and measures scintillation light detected by APV photodiodes and associated charge preamplifiers~\cite{Wang:2012ep}.
Groups of 16 APD’s are connected via T-cards in the calorimeter.
The preamplifiers convert the light/charge generated by the APV’s into voltage steps with a common gain of 1mV/fC.
The discharge time of the voltage steps is hardwired to $100\mu$s, hence the average single channel rate is limited to $10$kHz.
The step voltage signals get transmitted over cables to shapers in the FEE  cards  which split the signals into 2 channels of gain ratio 1/16, each followed by one 10 bit  digitizer channel \cite{HansPHOSmanual}.
The shapers convert the step signals into semi-gaussian pulse-shape (Fig. 6) which can be  fitted with a Gamma-2 timing function.
\begin{figure}[t!] 
  \centering 
 \includegraphics[width=0.75\linewidth]{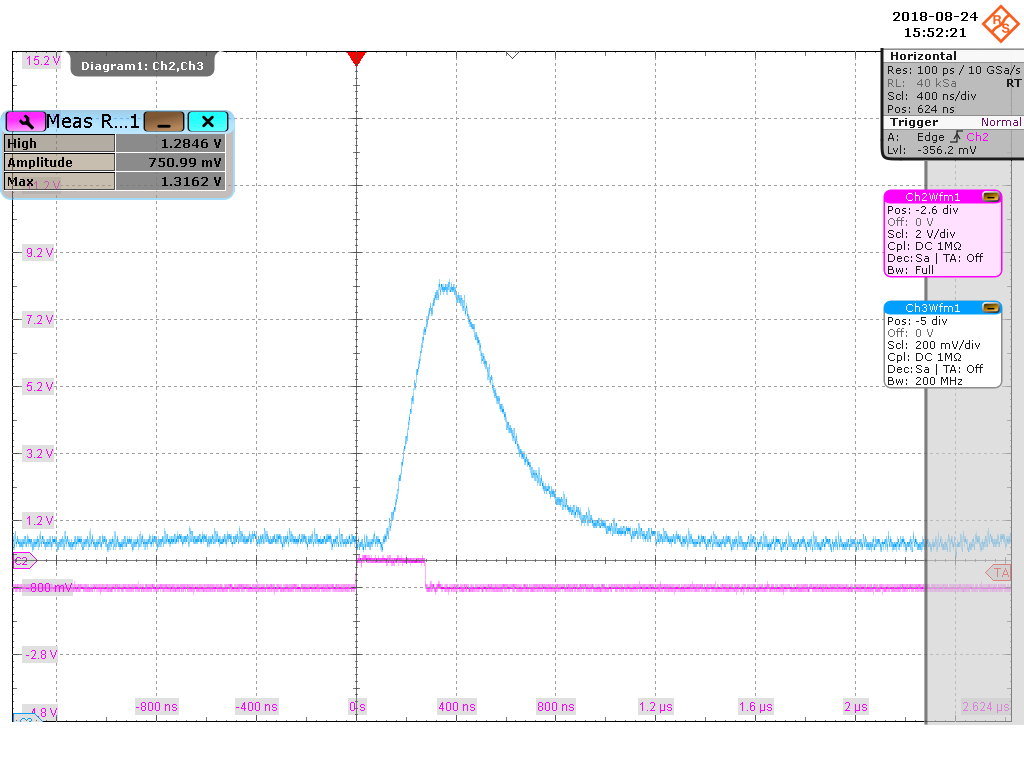}
 \caption{top: semi-gaussian shaper response to LED.   bottom: ALED trigger pulse.}
 \label{fig:6}
\end{figure}

\section{Dynamic range scan}
\begin{figure}[t!] 
  \centering 
 \includegraphics[width=0.75\linewidth]{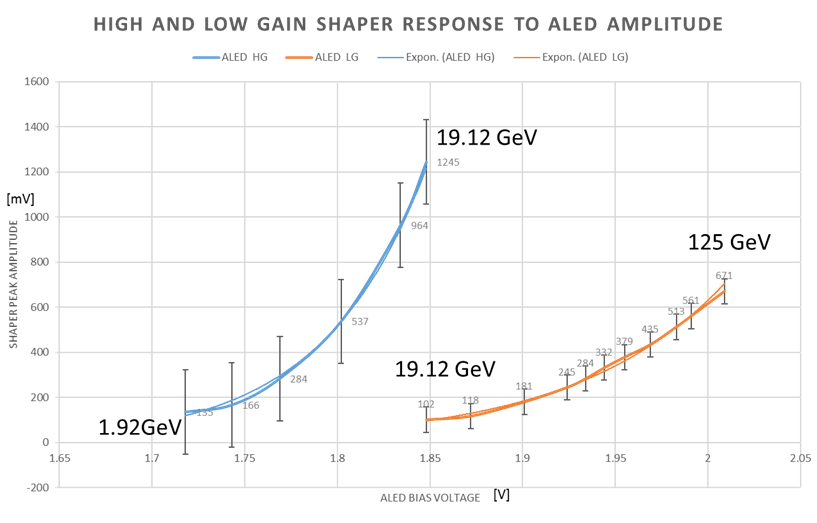}
 \caption{Dynamic range scan with LED pulser.}
 \label{fig:7}
\end{figure}

The electronics response to ALED pulsing was measured with a digital oscilloscope as an amplitude scan with the APD-preamplifier-FEE-shaper chain.
The peak amplitude response over the LED bias range is depicted in Fig.~\ref{fig:7}. 
The measurement reflects the exponential LED Diode forward characteristics over a range from $2$--$125$GeV, limited by noise below $2$GeV and by non-linear electronics response of the FEE electronics beyond $125$GeV. 
The equivalence of the peak shaper amplitude in Volt and the particle energy in GeV is determined over the FEE design parameters where the full range of the high gain  corresponds to $15$GeV for $1$V at the digitizer (Altro) input.
Therefore at high gain:
\begin{verbatim}
10 bit := 15000 MeV
1 bit =  1V/1024 = 0.9765 mV
E[MeV] = 15 * (Upeak /0.9765)
\end{verbatim}
Using the same equivalence and giving the measured high-gain maxiumum energy point (19.12 GeV in Fig.~\ref{fig:7}) the same energy for the lowest point of the low gain range, the energy equivalence for the low gain is 
$E[{\rm GeV}] = 0.183 * ({\rm Upeak} /0.9765)$ 
This equivalence is an approximation of the calibrated energy equivalence.
The results obtained here demonstrate  that with the  ALED pulser, the EMCal electronics chain reproduces the exponential LED forward law  $A \sim \exp ({\rm const} \cdot {\rm Ubias})$  over  the full dynamic range.

\section{Stability}
\begin{figure}[t!] 
  \centering 
 \includegraphics[width=0.49\linewidth]{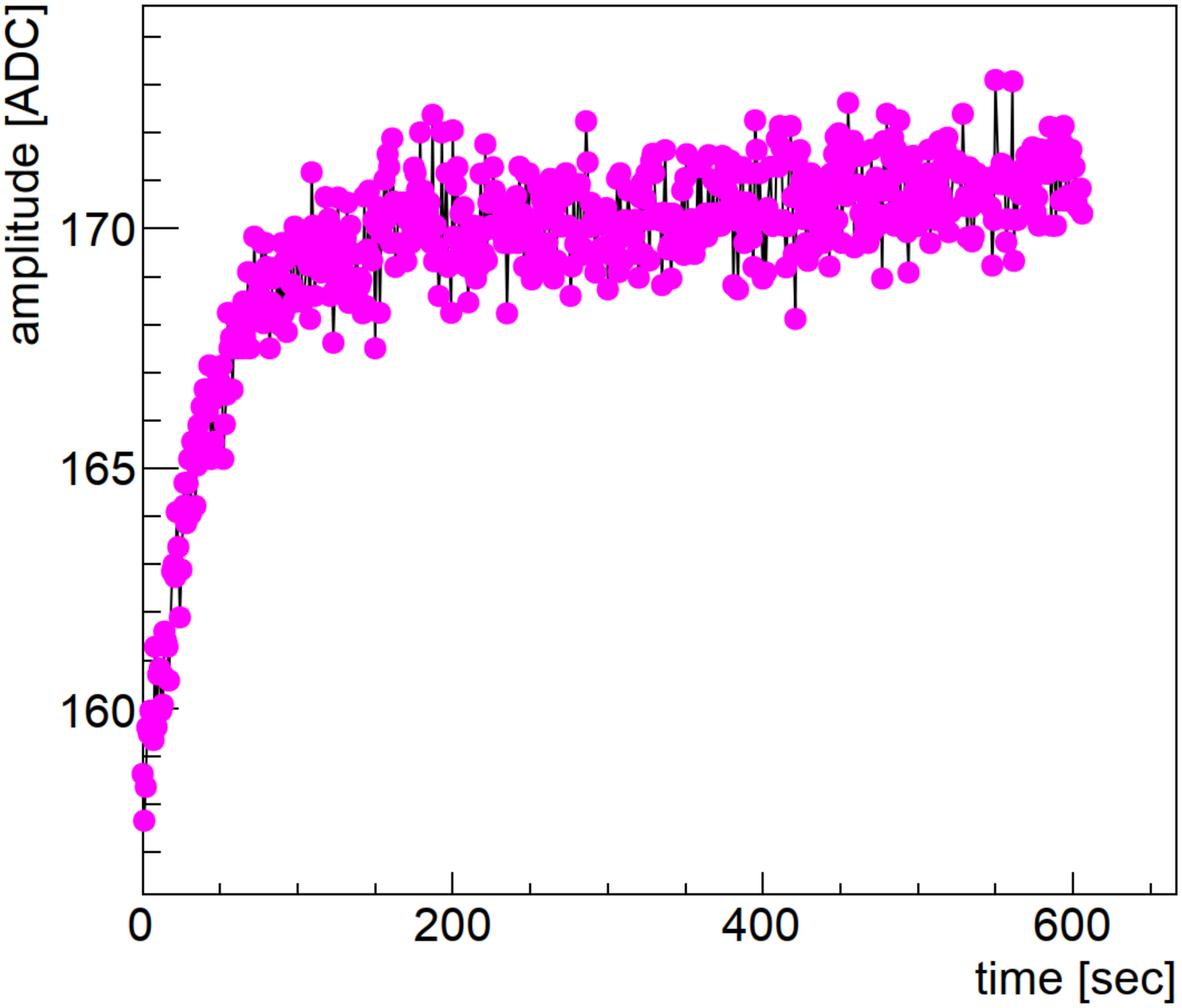}
 \includegraphics[width=0.49\linewidth]{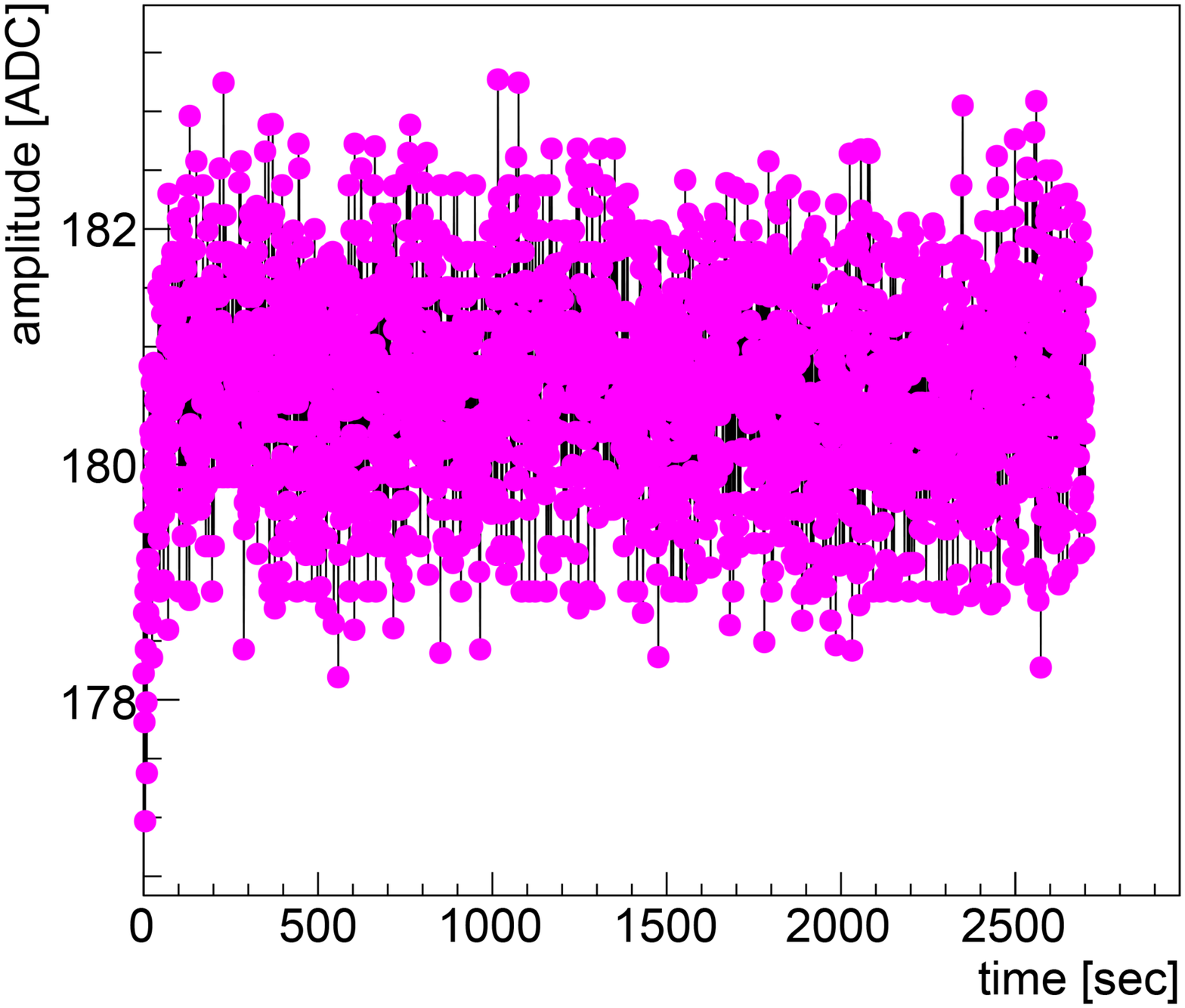}
 \caption{LED run amplitude over time.}
 \label{fig:8}
\end{figure}

The stability of the measured amplitude depends on the HV bias voltage, the APD temperature and the pulsed LED temperature.
Whilst the APD temperature and  the bias voltage can be well controlled, the temperature of the LED depends on the flashing frequency and on the bias voltage since it defines the amount of  generated light. 
As shown in the left panel of Fig.~\ref{fig:8}, 
the warming-up takes tens of seconds and the amplitude becomes stable after $\sim$100 $s$ of operation. 
As soon as the pulse amplitude reaches the maximum value, it remains stable with better than 0.5 $\%$ precision as shown on the right panel of Fig.~\ref{fig:8}. 




\section{Determination of saturation}
\begin{figure}[t!] 
  \centering 
 \includegraphics[width=0.75\linewidth]{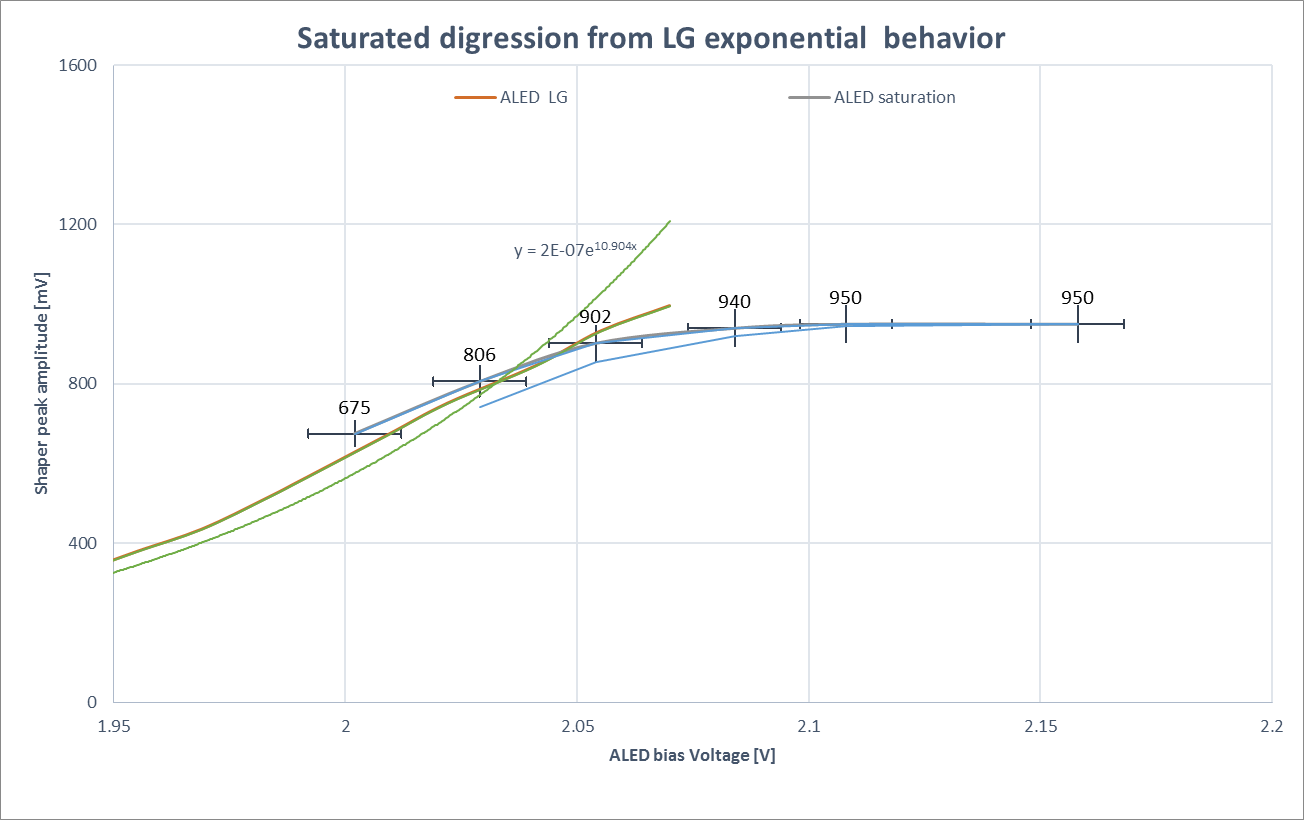}
 \caption{Low-gain saturation EMCal}
 \label{fig:9}
\end{figure}

The large dynamic range of ALED light was used to determine the limits of EMCal electronics linearity,  characterized by saturation of the electronics chain response to the exponential increase of the light  amplitude. This behaviour can be observed in Fig.~\ref{fig:9}.
\section{Crosstalk}
\begin{figure}[t!] 
 \centering 
 \includegraphics[width=0.75\linewidth]{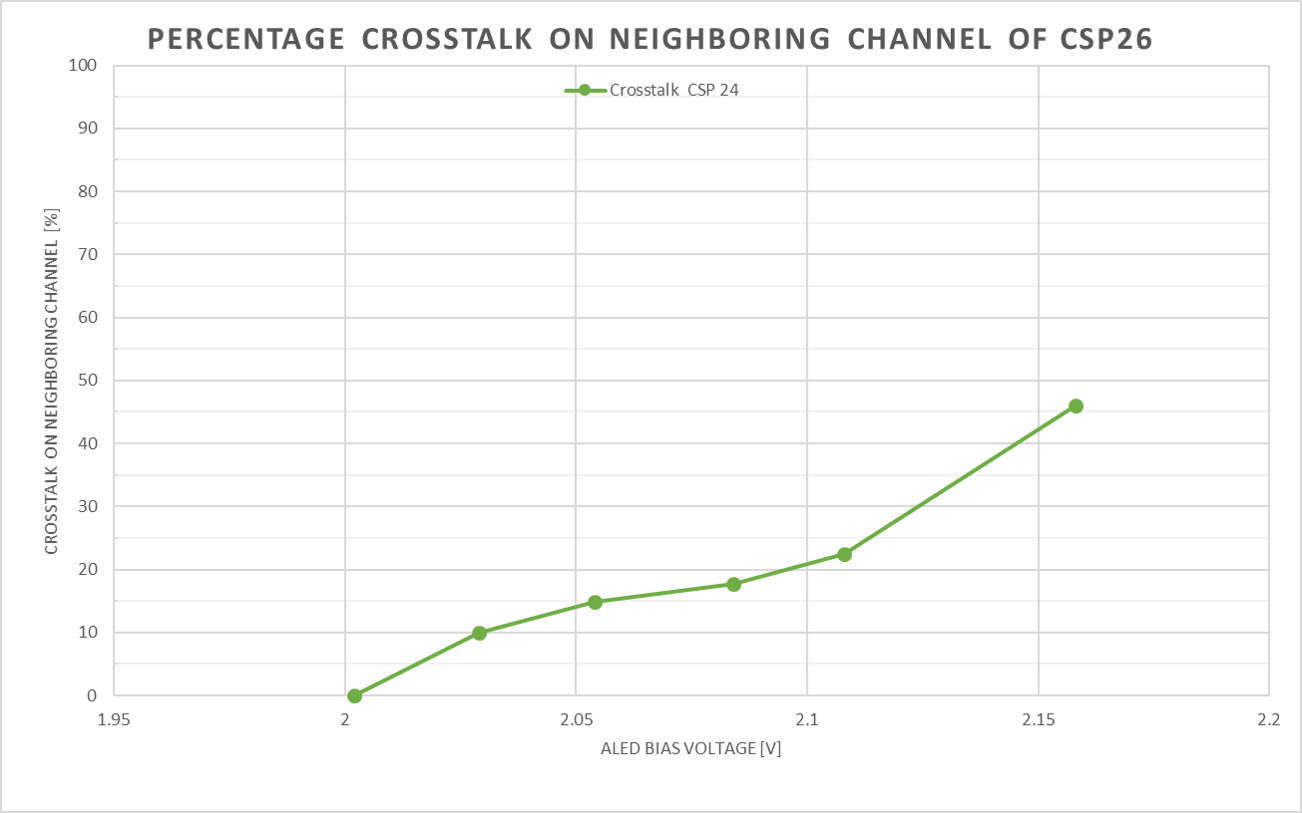}
 \caption{Crosstalk on T-card neighbor}
 \label{fig:10}
\end{figure}

The percentage of the signal crosstalk generated by a single pulsed channel to a neighboring channel on the T-card is shown as example in Fig.~\ref{fig:10}.
This measurement is however only exemplariy since the absolute crosstalk of particular channels depends strongly on the channel routing via cables and connectors and the PCB geometry between individual preamplifiers. 
A comprehensive crosstalk measurement therefore requires scanning of individually pulsed  channels and measuring  crosstalk  on all other channels.
A pulse corresponding to energy $E=100$~GeV is injected in the channel indicated with $\rm{col}=0$ and $\rm{row}=4$.
Blue histogram stands for High Gain and the red one for Low Gain.
One T-card is connected to 16(=2x8) channels and the time evolution of HG of other 15 channels are also shown.
The baseline variation of these channels indicates that a pulse is also induced in them and it is strongly correlated with the input pulse.
A detailed study showed that the cross-talk is mostly due to the ribbon cable connecting the T-card with FEC.
\begin{figure}[t!] 
 \centering 
 \includegraphics[width=0.9\linewidth]{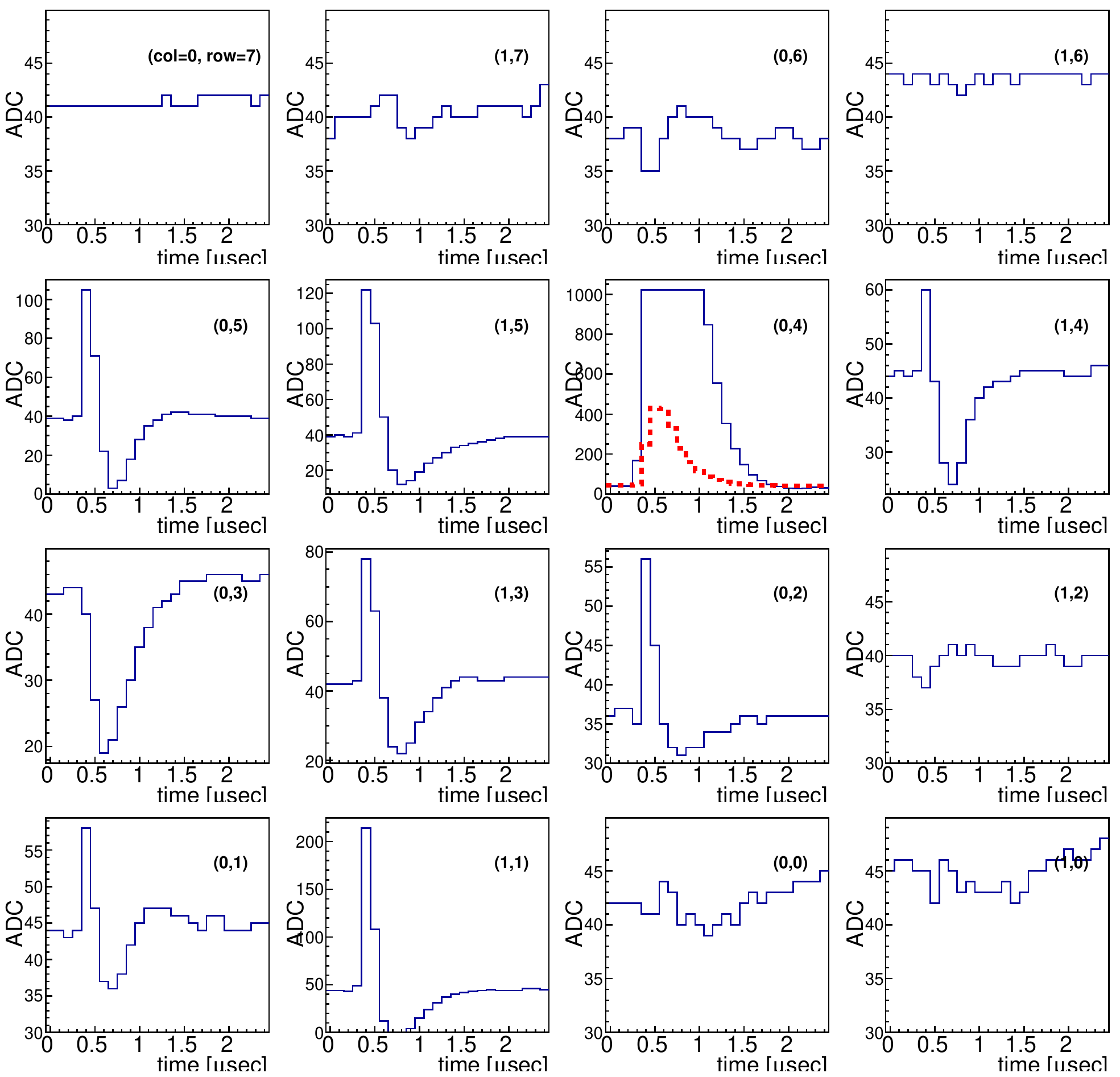}
 \caption{Crosstalk measurement of FEE neighbouring channels }
 \label{fig:11}
\end{figure}
\section{Summary}
A cost effective, novel LED pulser with eight simultaneously triggered, individual light pulse channels of tunable intensity was prototyped and tested with the FEE electronics of the ALICE EMCAL calorimeter.
The ALED generates light in a similar ways as scintillators with a sub-ns risetime, a few ns of high intensity, and followed by a decay time of tens of ns.
The light intensities of each individual channel can be tuned over the full dynamic range.
Due to absence of electrical crosstalk, non-linearities and crosstalk effects can be quantified more reliably than with previous electronic pulsers.
Multi-channel energy clustering can be emulated with proper energy determination of the analysis chain.    
\bibliographystyle{utphys}
\bibliography{aledpap}
\end{document}